\documentclass{ifacconf}

\usepackage{graphicx}      % include this line if your document contains figures
\usepackage{natbib}        % required for bibliography
\usepackage{amsmath,amssymb,amsfonts}
\usepackage{subfig}
\usepackage{multirow}

\newtheorem{remark}{Remark}
\begin{document}
\begin{frontmatter}

\title{Speed-Adaptive Model-Free Lateral Control for Automated Cars\thanksref{footnoteinfo}} 

\thanks[footnoteinfo]{This work is partially supported by the joint CNRS (France)-CSIC (Spain) PICS Program through the project UbiMFC (CoopIntEer 203 073)}%

\author[First]{Marcos Moreno-Gonzalez} 
\author[First]{Antonio Artuñedo} 
\author[First]{Jorge Villagra}
\author[Second,Fourth]{Cédric Join}
\author[Third,Fourth]{Michel Fliess}

\address[First]{Centre for Automation and Robotics (CSIC-UPM), ctra. de Campo Real, km 0,200, 28500 Arganda del Rey, Spain, (e-mail: \tt\small \{marcos.moreno, antonio.artunedo, jorge.villagra\}@csic.es).}
\address[Second]{CRAN (CNRS, UMR 7039), Universitée de Lorraine, BP 239, 54506 Vandoeuvre-lès-Nancy, France}
\address[Third]{LIX (CNRS, UMR 7161), École Polytechnique, 91128 Palaiseau, France, (e-mail: \tt\small Michel.Fliess@polytechnique.edu)}
\address[Fourth]{A.L.I.E.N., 7 rue Maurice Barrès, 54330 Vézelise, France, (e-mail: \tt\small \{cedric.join, michel.fliess\}@alien-sas.com)}

\begin{abstract}                % Abstract of not more than 250 words.
In order to increase the number of situations in which an intelligent vehicle can operate without human intervention, lateral control is required to accurately guide it in a reference trajectory regardless of the shape of the road or the longitudinal speed. Some studies address this problem by tuning a controller for low and high speeds and including an output adaptation law. In this paper, a strategy framed in the Model-Free Control paradigm is presented to laterally control the vehicle over a wide speed range. Tracking quality, system stability and passenger comfort are thoroughly analyzed and compared to similar control structures. The results obtained both in simulation and with a real vehicle show that the developed strategy tracks a large number of trajectories with high degree of accuracy, safety and comfort.
\end{abstract}


\end{frontmatter}
%===============================================================================

\section{Introduction}

Over the last years, autonomous driving capabilities have been developed, but in order to increase the situations in which the vehicle operates without driver intervention, accurately controlling the vehicle in any scenario is essential.

In a decoupled control architecture, lateral control keeps the vehicle on the path without losing stability or impairing passenger comfort, regardless of the road or speed. This problem has been addressed with different approaches, the first being to improve the available model of the vehicle in order to better tune a model-based regulator. Another is the design of controllers for different longitudinal speeds, and then provide a fitting law between their different parameters or outputs.  To avoid the problems induced by a complex system identification, some approaches opted to rely on model-independent strategies.

In this paper, a new lateral control strategy for autonomous vehicles is presented within the Model-Free Control (MFC) paradigm \citep{fliess.ijc13}. The proposed control law is based on the adaptation of one of the key parameters of MFC as a function of driving speed, allowing thus the vehicle to cope with a variety of situations without having to re-tune the controller. To thoroughly evaluate the potential of this strategy, metrics of tracking quality, stability of the feedback system and passenger comfort are defined. A multi-objective optimization has been applied in simulation to determine which control structure provides the best trade-off among the three criteria in a wide spectrum of situations. The main contribution of the work relies on the introduction of an easy-to-implement variation of MFC, which has proven to be very effective for lateral control of automated vehicles.

The rest of the paper is structured as follows. Section II presents a brief review of the lateral control strategies in the literature. A theoretical introduction to Model-Free Control is presented in Section III. Section IV motivates the proposed Speed-Adaptive Model-Free Control strategy. The results from simulation and real-world tests are presented in Section V. Finally, the last section draws some concluding remarks and the references.

\section{State of the art}

Lateral control of autonomous vehicles is studied from different approaches, most of them based on a somewhat realistic model of the vehicle, being the single-track model the most popular \citep{Arifin:icecos19}, which is linear and assumes a constant longitudinal velocity. Some real applications show that this simplification may be inappropriate to control the vehicle in any scenario. To overcome these limitations, different strategies have been proposed in the last years. \cite{zainal2017yaw} applied the single-track model to fit two PID regulators, one for low and one for high speeds; \cite{s18082544} used it to synthesize a robust LQR. Alternatively,  \cite{Zanon2014} rely on a more complex model to develop a Model Predictive Control (MPC) strategy, and \cite{8814173} focus on jointly solving the path planning and control problems; but this strategies are mainly tested on specific situations.

Real vehicles have complex dynamics that vary with speed and steering angle, with strong non-linearities, couplings between lateral and longitudinal dynamics and variability of parameters that are already difficult to characterize; consequently, it is extremely hard to find a realistic model for a large spectrum of driving situations. As a result, the potential of control strategies that do not rely on a vehicle dynamic model has catched attention. 

Fuzzy control is a good example of these model-free techniques, as it absorbs some of the variability of the system parameters and its formulation is intuitive, but difficult to tune optimally over a wide working range. Two fuzzy regulators were integrated and validated in traffic-based driving environments in \cite{godoy2015adriverless}; other works \citep{7963822} confirmed the capabilities of fuzzy logic for lateral control. Another approach is pure pursuit control \citep{6987787}, which is based on a kinematic model of the vehicle, but its performance degrades when high velocities or accelerations are requested. 

The MFC framework evoked in the introduction was successfully applied in vehicle longitudinal control \citep{villagra.ijvas09} or in lateral control for low-speed AGVs \citep{villagra.tcst12}. Alternatively, in \cite{6760313} the flatness theory \citep{fliess1995}, which allows finding differentially flat outputs for non-linear systems, is applied to implement the lateral control of a vehicle together with a model-free feedback controller. This approach exhibited very good performance in simulation, but its deployment in real vehicles requires measurements that cannot be obtained with commercial sensors. Alternatively, \citep{wang2022extremum} proposes an adaptation mechanism for MFC and apply it on a scale car, but the resulting adaptation dynamics is too slow for automated vehicles driving on real roads.

 %Ultra-local Model Predictive Control (ULMPC) \citep{WANG2020104482} is an original combination of model-free and model-based approaches, relying on the application of an MPC regulator over the ultra-local model proposed by MFC. However, the computing power required in ULMPC, although lower than in MPC, is still not negligible.

\section{Model-free control principles}

\cite{fliess.ijc13} state that the system dynamics can be approximated by an ultra-local model 
\begin{equation}
y\textsuperscript{(n)} = F + \alpha \cdot u %\overset{n}{\dot{y}} sería la forma de poner la derivada enésima coherente con la notación de punto, pero es menos conocida
\label{eq_ultralocal}
\end{equation}
in which the linear relationship between the input $u$ and the nth derivative of the output $y$  is fitted by a variable $F$ that absorbs model errors and system disturbances, and where the ratio constant $\alpha$ is a design parameter.

The control loop is closed by an \textit{intelligent PID} controller, iPID controller (usually iP or iPD):
\begin{equation}
u = \frac{1}{\alpha}  \cdot \left(-F + y_r ^{(n)} + K_p \, e + K_i \int{e} + K_d \, \dot{e}\right)  
\label{eq2}
\end{equation}
where $u$ is the control action, suffix $r$ means reference, $e$ is the tracking error and $K_p$, $K_i$ and $K_d$ are the control parameters, emulating those of a PID controller. The term $F$ must be estimated in real time, for this purpose, it can be assumed to be the same between consecutive instants and can be estimated from \eqref{eq_ultralocal} as follows: 
\begin{equation}
\hat{F} (t_k) = \hat{y} ^{(n)} (t_k) - \alpha \cdot u (t_{k - 1}) 
\label{eq3}
\end{equation}
where $\hat{F}$ is the estimator of $F$, $t_k$ is the current instant and $\hat{y} ^{(n)}$ is the filtered nth derivative of $y$.

\begin{remark}
Note that the error dynamics derived from (\ref{eq_ultralocal}) and (\ref{eq2}) can be expressed as  $f(e, K_p, K_i, K_d)=\hat{F}-F$. If the estimation of $F$ is good enough ($\hat{F} \approx F$), then the system dynamics could be made asymptotically stable through an appropriate choice of the control parameters.
\end{remark}
\section{Speed-adaptive lateral control}

The parameter $\alpha$ defines in a certain way the aggressiveness of the iP(D) controller, since the higher is $\alpha$, the smaller the increase in the control action between sampling instants. Therefore, varying $\alpha$ might adapt the controller aggressiveness to different driving situations.

\begin{figure}[htbp]
\centering
    \subfloat[Trajectory reference and tracking]{
        \includegraphics[width=0.97\linewidth]{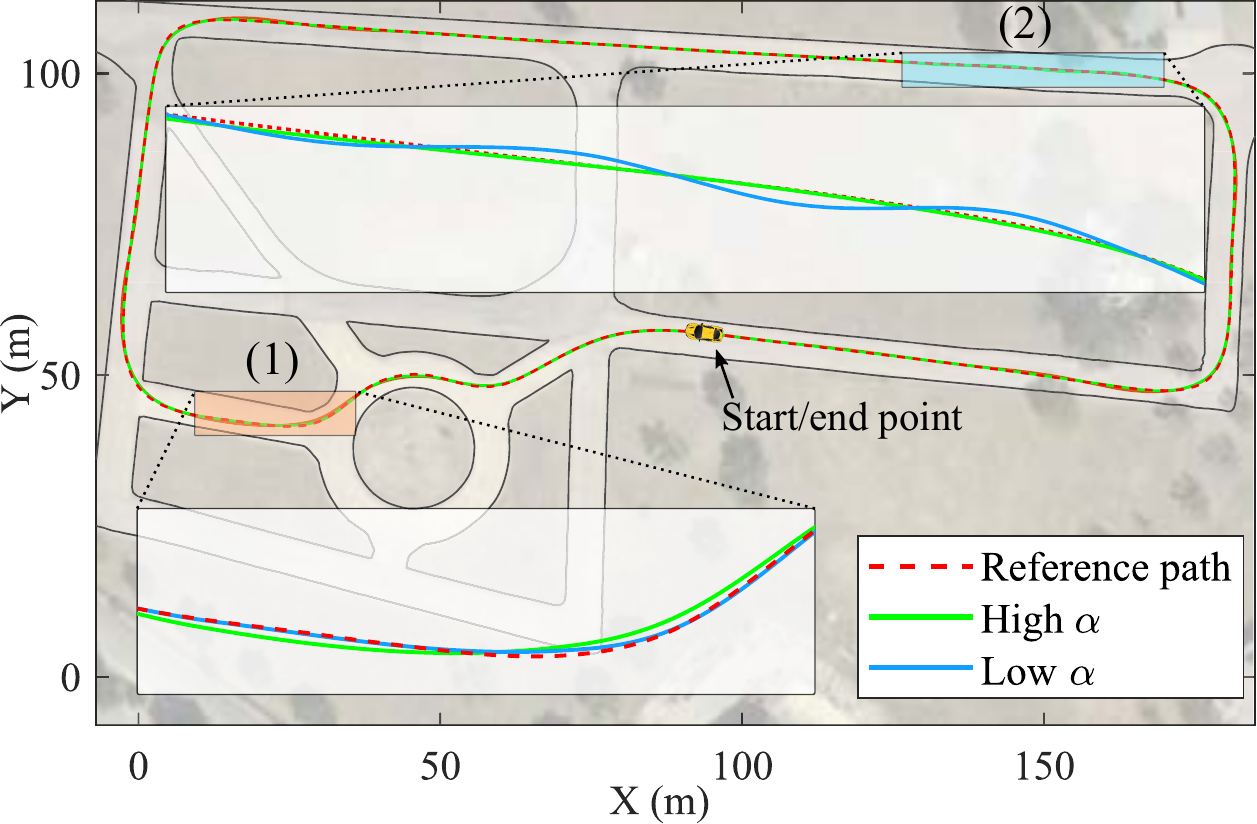}
        \label{fig_alfa_var:alfavar1}}
    
    \subfloat[Reference and vehicle speed]{
        \includegraphics[width=0.47\linewidth]{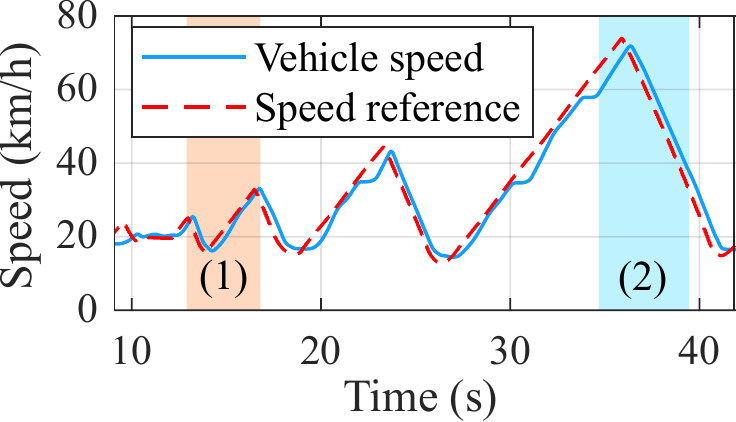}
        \label{fig_alfa_var:alfaalto2}}
    \subfloat[Steering wheel angle in (2)]{
        \includegraphics[width=0.47\linewidth]{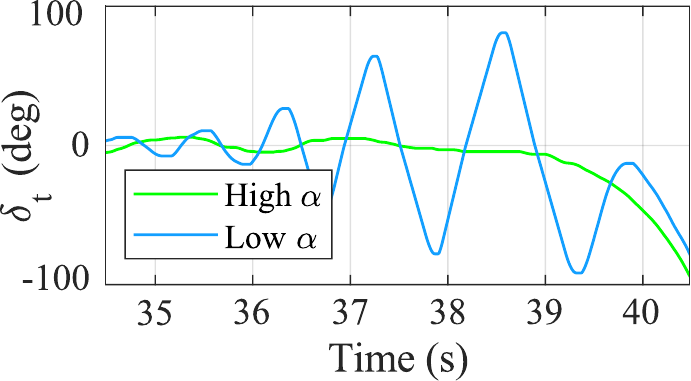}
        \label{fig_alfa_var:deltaT}}
\caption{Comparison between high and low alpha in low speed curves and higher speed straight stretches}
\label{fig_alfa_var}
\end{figure}

Fig. \ref{fig_alfa_var} shows the same MFC regulator with two different $\alpha$: the one with a low value performs better at low speed curves but becomes highly oscillating at a stretch where higher speed is allowed; the configuration with high $\alpha$ is stable for the straight path (with little oscillation) but does worse tracking at curves. This finding motivated the introduction of a model-free controller whose $\alpha$ varies as a function of speed $v$. This controller has a base $\alpha _0$ which is kept up to a given speed $v_{0}$, after which it is increased proportionally to speed variation with a constant $K_\alpha$:
\begin{equation}
\alpha = \max \left\lbrace\alpha _{0}, \, K_\alpha \cdot (v - v_{0}) + \alpha _{0}\right\rbrace
%    \begin{cases}
%        \alpha _{0} & \text{if}\; v < v_{0} \\
%        K_\alpha \cdot (v - v_{0}) + \alpha _{0} & \text{if}\; v \geq v _{0}
%    \end{cases}
    \label{eq6}
\end{equation}
This $\alpha$ variation law allows to obtain a (i) more aggressive behaviour in urban environments and (ii) smoother actions on the highway, where the oscillations can impair comfort and lead to system instability due to the high speed. 

\section{Results}

In this section, the control parameter space is explored in simulation (section \ref{section:exploration}) using a high-fidelity vehicle model (section \ref{section:model}). This evaluation is done with respect to three metrics (defined in \ref{section:metrics}) and is applied to the proposed feedback control approach, a standard MFC and a PID in a wide set of driving contexts (described in section \ref{section:benchmark}). To confirm the suitability of the estimated performance potential for each controller (obtained in \ref{section:sim_results}), experimental trials have been prepared and conducted in an automated prototype, and are reported in section \ref{section:experiments}.

\subsection{Benchmark description}\label{section:benchmark}

\begin{figure*}[htbp]
\centering
    \subfloat[Trajectory $T_1$]{
        \includegraphics[width=0.37\linewidth]{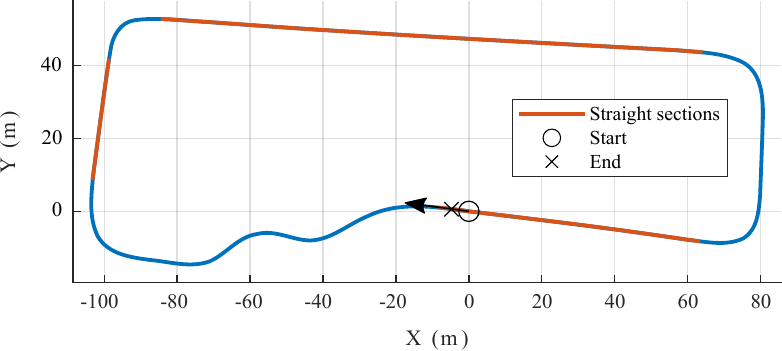}
        \label{fig:T1}}
    \subfloat[Trajectory $T_2$]{
        \includegraphics[width=0.325\linewidth]{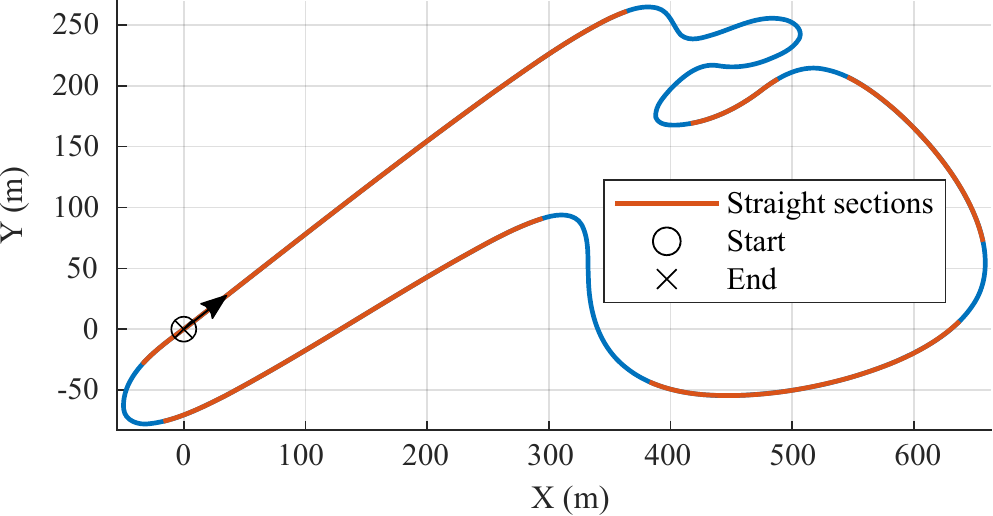}
        \label{fig:T2}}
    \subfloat[Trajectory $T_3$]{
        \includegraphics[width=0.24\linewidth]{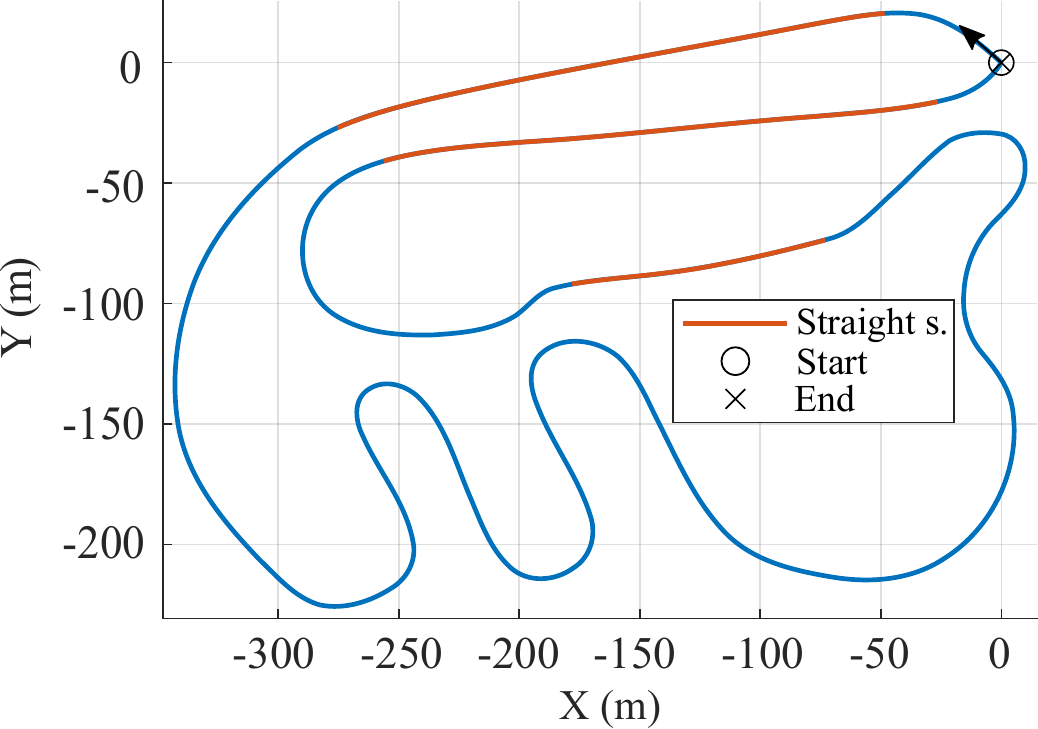}
        \label{fig:T3}}
\caption{Benchmark trajectories}
\label{fig:Ts}
\end{figure*}

Fig. \ref{fig:Ts} shows an aerial view of the three circuits used in the simulations, which are designed to cover a wide range of realistic driving scenarios. They have straight sections, wide and sharp curves and different dynamic constraints, as can be seen in Table \ref{tab:trajectories}. The reference trajectories are generated for each path applying the method for acceleration-limited speed planning proposed in \cite{Artunedo2021}.

%
%\begin{figure*}%[ht!]
%    \centering
%    \includegraphics[width=1\linewidth]{figures/all_lateral_error_dc.pdf}
%    \caption{Lateral error of each controller}
%    \label{fig_results_lateral_error}
%\end{figure*}

\begin{table}[htpb]
\centering
\captionsetup{width=\linewidth}
\caption{Dynamic constraints of each trajectory}
\label{tab:my-table}
\begin{tabular}{|l|l|l|l|}
\hline
Trajectory                       & $T_1$  & $T_2$  & $T_3$  \\ \hline
Maximum speed ($km/h$)               & 35  & 100 & 70  \\ \hline
Maximum long. acceleration ($m/s^2$) & 0.4 & 1.5 & 2.0 \\ \hline
Maximum long. deceleration ($m/s^2$) & 0.7 & 2.0 & 2.0 \\ \hline
Maximum lat. acceleration ($m/s^2$)  & 1.0 & 4.0 & 2.0 \\ \hline
\end{tabular}
\label{tab:trajectories}
\end{table}

\subsection{Metrics description}\label{section:metrics}

The integral absolute lateral error (IAE) is used to evaluate the tracking quality. But it is observed that the classical IAU (Integral of the Absolute control action) is not suitable for analyzing the system dynamics, the frequency spectrum of the feedback control action is used to define two different performance indicators:

\begin{enumerate}
    \item $M_\epsilon$ quantifies the low frequency oscillations of the control action, which can lead to vehicle instability, as well as impair the comfort of the occupants.

    \item $M_\zeta$ quantifies the high frequency oscillations of the control action, which do not destabilize the system per se, but cause discomfort to the vehicle occupants.
\end{enumerate}

The values of both metrics, $M_\epsilon$ and $M_\zeta$, are computed from two separated frequency bands, experimentally identified: $\epsilon$ (1.1-4 Hz) and $\zeta$ (4-10 Hz), respectively. Note that 20 Hz is the control frequency in this work. A high-pass filter is firstly applied to the feedback control action with a cutoff frequency of 0.5 Hz and 4 Hz respectively. Then, the spectrum is calculated by applying the Short-time Fourier transform with 5-second overlapping sections.

$M_\epsilon$ is finally calculated as the mean of the maximum power spectrum at each section, considering a threshold and a scale factor to balance the magnitude of both metrics:
\begin{equation*}
    M_\epsilon = \frac{1}{n}\sum_{i=1}^{n} s_\epsilon \cdot \max{(10\cdot\log{P_{\epsilon,i}}+\lambda_\epsilon)}
    \label{eq:m_eps}
\end{equation*}
where $n$ is the amount of 5-second sections, $P_{\epsilon,i}$ is the spectrum power of band $\epsilon$ in section $i$, $s_\epsilon$ is a scale factor ($s_\epsilon=0.015$) and $\lambda_\epsilon$ is a threshold in dB ($\lambda_\epsilon = 80 dB$). Note that only straight and long sections where the path curvature is below $0.01$  m$^{-1}$ (drawn in orange in Fig.~\ref{fig:Ts}) are considered due to the high sensitivity of $M_\epsilon$.

$M_\zeta$ is obtained with the maximum (instead of the mean) power among all sections, using a scale factor of $s_\zeta=0.04$ and the same threshold $\lambda_\zeta = 80 dB$. This particularity is motivated by the low equivalence found in experimental tests between what is intended to be reflected by this metric and the value obtained with the mean. However, when the maximum power is considered, controllers that exhibit high frequency oscillations in any section of the test trajectory are penalised with high values of $M_\zeta$.

To summarise, IAE is the chosen indicator of the reference tracking quality, $M_\epsilon$ measures the (in)stability margin of the controller and $M_\zeta$ reflects passenger discomfort.

\subsection{Control parameter space exploration}\label{section:exploration}

Design trade-offs between tracking accuracy and control action safety and softness depend on driving dynamics. It is therefore difficult to determine a unique benchmark to assess the performance of the controller. Besides, it is also hard to infer the control parameters of a given control structure for a precise value of a metric. To overcome this, a multi-objective optimization problem is defined:% as $\underset{\mathbf{\dot{x}}=\mathbf{f}(\mathbf{x},\,\mathbf{p}),\,\mathbf{p}\in[\mathbf{p}_{min},\,\mathbf{p}_{max}]}{\min} \;\mathbf{J}(\mathbf{p})$
\begin{equation*}
  \underset{\mathbf{\dot{x}}=\mathbf{f}(\mathbf{x},\,\mathbf{p}),\,\mathbf{p}\in[\mathbf{p}_{min},\,\mathbf{p}_{max}]}{\min} \;\mathbf{J}(\mathbf{p})
\end{equation*}
where the set of objective functions is $\mathbf{J(p)}=(\max($IAE$_i),$ $\, \max(M_{\epsilon,i}),\,\max(M_{\zeta,i}))^T$, $i=1..3$ is the number of circuit $T_i$ and $\mathbf{p}$ is the set of control parameters that will be compared in Section \ref{section:sim_results}. Note that the functions to minimize are constrained by the system dynamics and the control parameters are bounded by design.

To solve this problem, the MATLAB's ParetoSearch algorithm is used. A Pareto front is obtained for each controller showing its potential to minimize each objective. With the maximum acceptable values for the optimization objectives and the Pareto front of each controller, a region is defined. The smaller the region, the greater the potential of a controller; thus, several controllers can be compared, even if they have very different characteristics. The acceptable optimization objective zone has been defined experimentally: It is observed that an IAE greater than 0.35 meters implies poor tracking in the curves; similarly, an $M_\epsilon$ greater than 0.25 implies the possibility of system instability on test trajectories; and a $M_\zeta$ greater than 0.7 leads to a loss of passenger comfort.

In order to make the results more general, every controller is simulated in the three benchmark trajectories $T_i$, which cover a wide operation range. The three quality metrics are obtained for each trajectory, and the maximum of each metric is considered in the optimization. It is thus ensured that the controller will not be less stable, comfortable or accurate in any situation.

\subsection{Vehicle model used in simulation}\label{section:model}

The model used in simulations mimics with a high degree of fidelity the experimental platform used, as it has 14 degrees of freedom (6 for the vehicle body motion: longitudinal, lateral, vertical, roll, pitch, and yaw; and 8 for the wheels: vertical motion and spin of each wheel). 

The power-train modeling comprises: (i) the engine, whose torque map is modeled from experimental measurements; (ii) the gearbox, which includes the same drive ratios and gear shifting logic as the real vehicle; (iii) the resistance torques coming from braking system and wind and gravitational forces.
The tire behaviour was reproduced with the Pacejka tire model~\citep{Pacejka1992}.

An external actuation system has been added to the vehicle's assisted steering, which is modeled as in~\cite{7886365}. Its main parameters, such as inertia or backlash, have been identified from extensive field tests.
Moreover, the small noise coming from the localization system has been characterized and included in the model.

\subsection{Simulation results}\label{section:sim_results}

All controllers implemented include a feedforward term $\delta_{ff}$ \citep{godoy2015adriverless}: $\delta_{ff} = R_S \cdot \arctan{\left(L \cdot \kappa\right)}$,
%
%\begin{equation}
% \delta_{ff} = R_S \cdot \arctan{\left(L \cdot \kappa\right)}
%\label{eq_feedforward}
%\end{equation}
%
where $L$ is the wheelbase, $\kappa$ is the path curvature and $R_S$ is the steering ratio. Note that this feed-forward term relies on a simplified kinematic vehicle model and the path curvature. 
The total steering control action is $\delta_t = \delta_{ff} + \delta_{max} \cdot u_{fb}$,
%
%\begin{equation}
%\delta_t = \delta_{ff} + \delta_{max} \cdot u_{fb}
%\end{equation}
%
where $u_{fb}\in[-1,1]$ is the normalized feedback control action and $\delta_{max}$ is the maximum steering angle.
%
%\begin{figure}[htbp]
%\centering
%\includegraphics[width=0.55\linewidth]{figures/dibujo.pdf}
%\caption{Kinematic model parameters}
%\label{fig_feedforward}
%\end{figure}

Three strategies have been evaluated: a discrete PID, an iPD (hereinafter MFC) and a Speed-Adaptive iPD (SAMFC). The two latter control laws use a second order ultra-local model \citep{fliess2021alternative}, so that the discrete form of \eqref{eq2} with no integral action yields:
\begin{eqnarray}\nonumber
u_{iPD} (t_k) &=& \frac{- \hat{F}(t_k) + \ddot{y}_{1r}(t_k) + K_p \, e(t_k) + K_d \, \hat{\dot{e}}(t_k)}{\alpha}\\\label{eq_iPD}
\hat{\dot{e}}(t_k) &=& \dot{y}_{1r}(t_k) - \hat{\dot{y}}_{1}(t_k)\\\nonumber
\hat{\dot{y}}_{1}(t_k) &=& \frac{ 2 y_{1}(t_k) - 2 y_{1}(t_{k-1}) - (T_s - 2 T_c)\hat{\dot{y}}_{1}(t_{k-1})}{T_s + 2 T_c}
\end{eqnarray}
where $y_1$ is the lateral deviation, $\hat{\dot{y}}_{1}$ is the estimated derivative of the vehicle's lateral deviation, $\hat{\dot{e}}$ is the filtered estimation of the tracking error derivative, $T_s$ is the sample time and $T_c$ is the derivative filter parameter.

Note that $\alpha$, $K_p$ and $K_d$ are the parameters that are varied in the optimization of the MFC controller. For SAMFC, the constant $\alpha$ is replaced by \eqref{eq6}, so that $\alpha_0$, $K_\alpha$  and $v_0$ are included in the design parameter set.  

The PID controller is expressed as a Z-transform function
\begin{equation}
U_{PID}(z) =  \left(K_p + \frac{K_i \cdot T_s}{z-1} + \frac{K_d \cdot N}{1+N \cdot T_s \frac{1}{z-1}}\right)E(z)
\label{eq_PID}
\end{equation}
being $K_p$, $K_i$, $K_d$ gains and filter parameter coefficient $N$ the parameters varied in the multi-objective optimization.

\begin{figure}[htbp]
\centering
    \subfloat[3D view]{
        \includegraphics[width=0.94\linewidth]{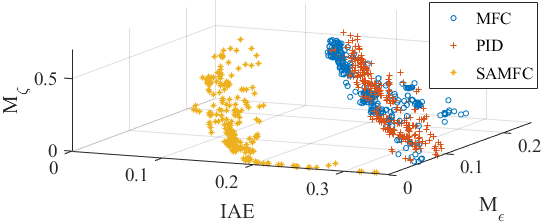}
        \label{fig_pareto:pareto3D}}
        
    \subfloat[IAE vs. M$_\epsilon$]{
        \includegraphics[width=0.94\linewidth]{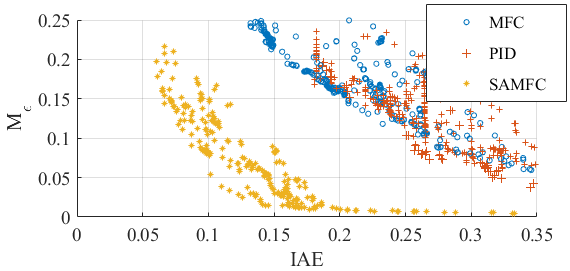}
        \label{fig_pareto:IAE_FM1}}
        
    \subfloat[IAE vs. M$_\zeta$]{
        \includegraphics[width=0.94\linewidth]{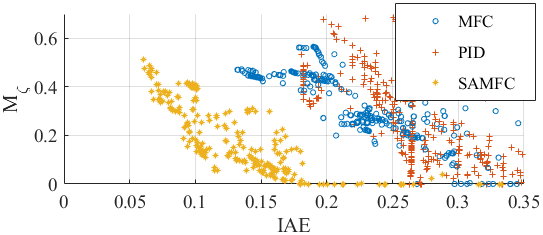}
        \label{fig_pareto:IAE_FM2}}
\caption{3D Pareto front for MFC, PID and SAMFC}
\label{fig_pareto}
\end{figure}

Fig. \ref{fig_pareto} shows the resulting Pareto fronts, including a 3D view and its projections on the IAE-$M_\epsilon$ and IAE-$M_\zeta$ planes. Remark that while the Pareto front of the basic MFC regulator is very similar to the front of the PID controller, the front of the SAMFC shows a significant reduction on the three metrics at the same time. Fig. \ref{fig_pareto} also shows that the Pareto fronts of the tested MFC and PID intersect and intersperse within the operation range, making it difficult to compare both strategies.

The SAMFC regulator is able to cancel $M_\epsilon$ within the operation range defined, as can be seen in Fig. \ref{fig_pareto:IAE_FM1}, in contrast to the MFC and PID regulators, whose minimum $M_\epsilon$ is 0.05 approximately. While inside the operation range MFC and PID have a minimum IAE of 0.13 and 0.17 meters respectively, the SAMFC regulator reach 0.06 meters. On the other hand, the three strategies offer several configurations where $M_\zeta$ is cancelled, as Fig. \ref{fig_pareto:IAE_FM2} shows, meaning the passenger comfort is assured.

Table \ref{tab:volumes} gathers the volume of the region defined by the Pareto front of the three strategies and the bounds of the operating region. As the points of the fronts are not equally distributed, the volume is obtained by numerical integration over the lineally interpolated surface between the points. The smaller the volume of the region (VUP), the higher the optimization potential of the controller structure, so, as can be seen in the table, SAMFC has a significantly better performance than its competitors.
\begin{table}
\centering
\captionsetup{width=\linewidth}
\caption{Volume under the Pareto front (VUP)}
\begin{tabular}{|l|c|c|c|} 
\hline
Control strategy & PID \eqref{eq_PID} & MFC \eqref{eq_iPD} & SAMFC \eqref{eq_iPD}{\scriptsize +}\eqref{eq6}  \\ 
\hline
VUP              & 0.0427      & 0.0404      & \textbf{0.0135}             \\
\hline
\end{tabular}
\label{tab:volumes}
\end{table}

\subsection{Experimental results}\label{section:experiments}

\begin{figure}[htpb]
    \centering
    \includegraphics[width=0.7\linewidth]{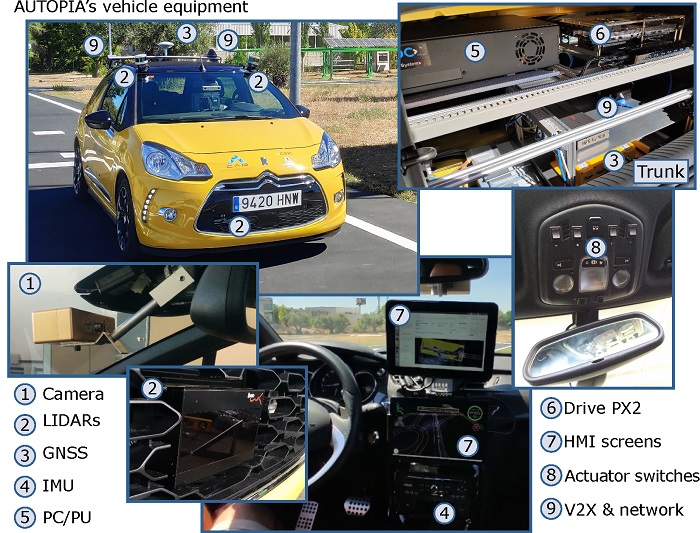}
    \caption{Experimental platform}
    \label{fig:ds3}
\end{figure}

The vehicle used in the experiments is a Citro\"{e}n DS3 which includes hardware modifications for the automated control of throttle, brake, gearbox and steering systems (see Fig.~\ref{fig:ds3}). Its localization relies on a RTK-DGPS receiver and on-board vehicle speed, accelerations and yaw rate sensors. An on-board computer with an Intel Core i7-3610QE and 8Gb RAM is used for control. It is also equipped with sensors to perceive the environment~\citep{8814070}.

From the acceptable operating region defined in section~\ref{section:exploration}, the parameter set that yields the best tracking quality is selected for each controller. The values are shown in Table \ref{tab:param}. Note that this decision yields for this specific system a MFC controller configuration with $K_p=0$.

\begin{table}[htpb]
\scriptsize
\centering
\captionsetup{width=\linewidth}
\caption{Parameters set for each controller}
\label{tab:param}
\tabcolsep=0.095cm
\begin{tabular}{|l|c|c|c|c|c|c|c|} 
\hline
Contr. & $K_p$     & $K_d$     & $N$ & ${\alpha} / {\alpha}_{0} $ & $K_{\alpha}$ & $v_0$  \\ 
\hline
PID        & 0.2216 & 0.0367 & 5.0 & -               & -         & -     \\ 
\hline
MFC        & 0      & 19.28  & - & 1409            & -         & -     \\ 
\hline
SAMFC      & 0.5625      & 2.688  & - & 57.15            & 9.547        & 26.83 \\
\hline
\end{tabular}
\end{table}

\begin{table}[htpb]
\centering
\captionsetup{width=\linewidth}
\caption{Setups tested in the experimental platform}
\begin{tabular}{|l|c|c|}
\hline
Trajectory                       & $S_1$  & $S_2$  \\ \hline
Max speed ($km/h$)               & 35  & 56 \\ \hline
Max long. acceleration ($m/s^2$) & 0.4 & 1.0 \\ \hline
Max long. deceleration ($m/s^2$) & 0.7 & 2.0 \\ \hline
Max lat. acceleration ($m/s^2$)  & 1.0 & 2.0 \\ \hline
\end{tabular}
\label{tab:setups}
\end{table}

The path of $T_1$ (see Fig.~\ref{fig:T1}) was used to generate two different testing trajectories using the constraints in Table~\ref{tab:setups}. Then, the controllers were evaluated on both trajectories.

\begin{table}
\centering
\captionsetup{width=\linewidth}
\caption{Results obtained for each controller and trajectory}
\label{tab:results}
\begin{tabular}{|c|c|c|c|c|c|c|} 
\hline
\multirow{2}{*}{\begin{tabular}[c]{@{}c@{}}\\Controller\end{tabular}} & \multicolumn{3}{c|}{$S_1$}                   & \multicolumn{3}{c|}{$S_2$}                    \\ 
\cline{2-7}
                                                                      & IAE            & $M_\epsilon$ & $M_\zeta$ & IAE            & $M_\epsilon$ & $M_\zeta$  \\ 
\hline\hline
PID      \eqref{eq_PID}                                                      & 0.16          & 0.002        & 0.17     & 0.15          & 0.132        & 0.79      \\ 
\hline
MFC      \eqref{eq_iPD}                                                     & 0.12          & 0.017        & 0.00     & 0.10          & 0.151        & 0.25      \\ 
\hline
\begin{tabular}[c]{@{}c@{}}SAMFC \eqref{eq_iPD}$+$\eqref{eq6}\end{tabular}           & \textbf{0.03} & 0.246        & 0.68     & \textbf{0.03} & 0.220        & 0.79      \\
\hline
\end{tabular}
\end{table}

\begin{figure}
    \centering
    \includegraphics[width=\linewidth]{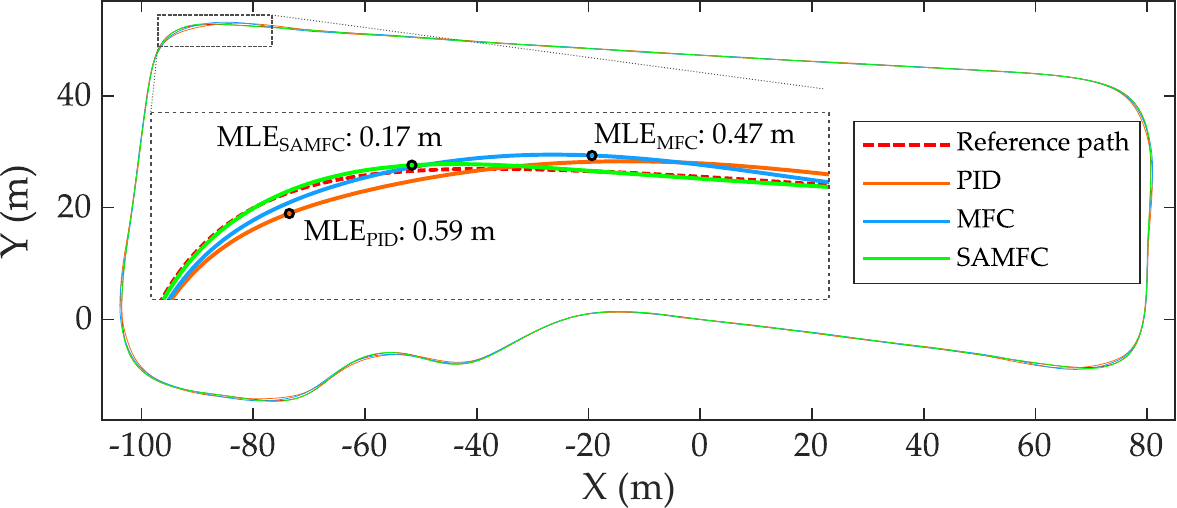}
    \caption{Tracking results on trajectory $S_1$}
    \label{fig:xy}
\end{figure}

\begin{figure*}%[ht!]
    \centering
    \includegraphics[width=\linewidth]{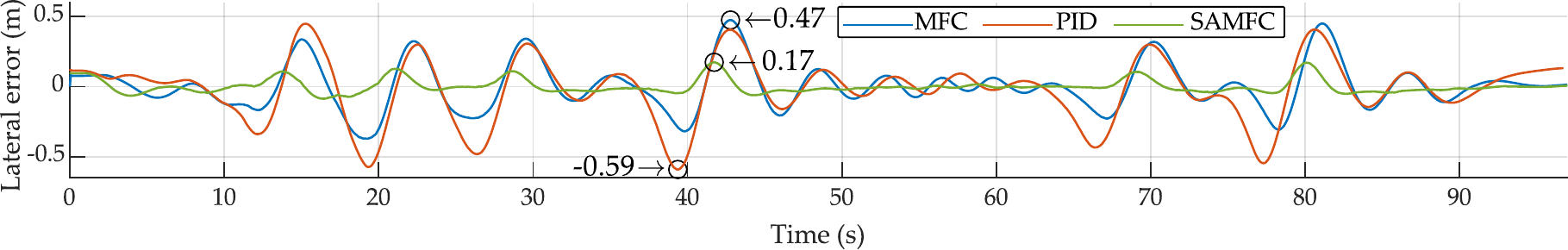}
    \caption{Lateral error of each controller}
    \label{fig_results_lateral_error}
\end{figure*}

\begin{figure}[htpb!]
\centering
    \subfloat[PID]{
        \centering \includegraphics[width=0.97\linewidth]{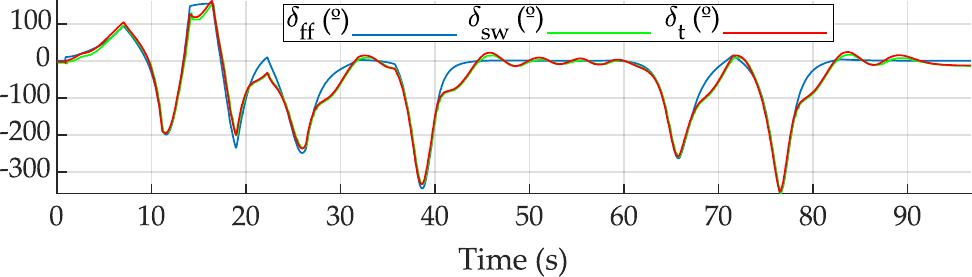}
        \label{fig_PID_ca}}

    \subfloat[MFC]{
        \centering \includegraphics[width=0.97\linewidth]{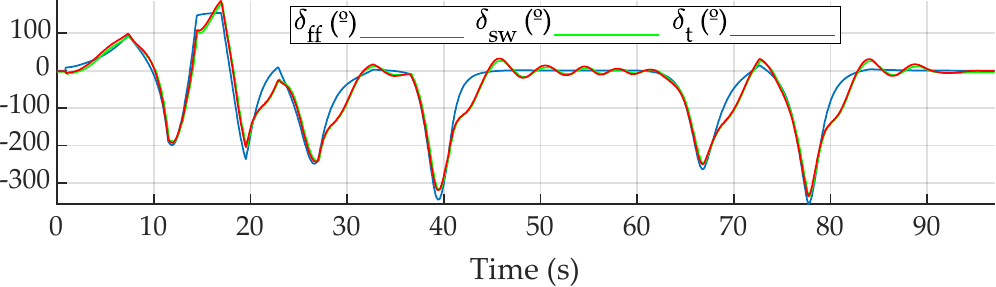}
        \label{fig_MFC_ca}}

    \subfloat[SAMFC]{
        \centering \includegraphics[width=0.97\linewidth]{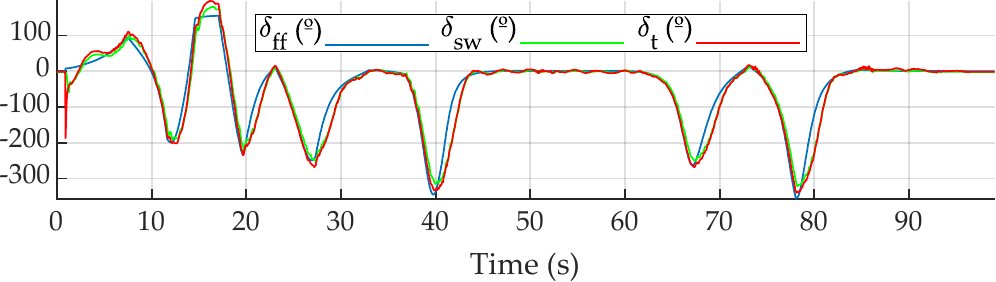}
        \label{fig_SAMFC_ca}}
\caption{Control actions of each controller}
\label{fig_results}
\end{figure}

The paths followed by the vehicle to track $S_1$ are depicted in Fig.~\ref{fig:xy}. As shown, all tested controllers are able to complete the circuit with small lateral errors. Nevertheless, some differences can be appreciated in the sharpest curves, where the Maximum Lateral Error (MLE) is significantly reduced by the SAMFC. In addition, Table~\ref{tab:results} shows clear differences in the tracking quality indicator (IAE, in meters): SAMFC obtains a reduction of more than 70\% with respect to MFC and PID. In order to analyze the performance in more detail, Fig.~\ref{fig_results} and Fig.~\ref{fig_results_lateral_error} shows the control action and the lateral error over time, respectively, where $\delta_{sw}$ the measured steering wheel angle.

None of the controllers exhibit uncomfortable or near-unstable behavior, as it is confirmed by $M_\epsilon$ and $M_\zeta$. They remain within the limits of the operating region, with the exception of $M_\zeta$ for SAMFC, that is slightly higher than expected due to minor unmodeled behaviours. However, the performance of SAMFC in the longest straight section of $S_1$ presents less low-frequency oscillations than MFC and PID, as can be observed between instants t = 47 s and t = 60 s in Fig.~\ref{fig_results}. Note that the SAMFC is able to achieve a very good tracking at low speeds (sharp turns) while limiting the amplitude of low frequency oscillations at higher speeds, resulting in a more stable behaviour.

Finally, it is worth highlighting that the adaptation mechanism of $\alpha$ is simple to implement and decoupled from the feedback gains, which makes the method easier to tune than a parameter varying PID.

\section*{Concluding remarks}

The aim of this work was to obtain a lateral control strategy with a wide operation range in order to increase the situations in which an autonomous vehicle navigates without intervention. A minimum level of safety and comfort were imposed. To that end, a systematic procedure to explore the controllers performance has been applied on trajectories with different shapes and dynamic constraints.

Speed-Adaptive MFC has demonstrated both in simulation and in real-world tests that it can outperform other controllers with a similar structure, such as a PID and a MFC. Furthermore, the changes introduced to the basic MFC do not significantly increase the controller tuning difficulty, which, together with the simple theoretical basis of MFC, results in an intuitive and efficient control strategy.

Future work will assess the tracking capabilities and robustness of SAMFC with respect to other model-based control strategies, such as Model Predictive Control.
%\printbibliography
%\bibliographystyle{plainnat}
% \bibliography{IEEEabrv,ifacconf}

\bibliography{ifacconf}             % bib file to produce the bibliography
                                                     % with bibtex (preferred)
                                                     
\end{document}